\documentclass[journal=jacsat,manuscript=article]{achemso}   %%% Original
\usepackage{amsfonts}
\usepackage{amssymb}
\usepackage{graphicx}
\usepackage{amsmath}
\usepackage{tikz}
\usepackage{adjustbox}
\usepackage[normalem]{ulem}
\usetikzlibrary{shapes.geometric, arrows}
\usepackage[english]{babel}
\usepackage{color}
\usepackage[version=3]{mhchem} % Formula subscripts using \ce{}
\usepackage{natbib}
\usepackage{url}
\usepackage[colorlinks,citecolor=black,urlcolor=blue,linkcolor=black]{hyperref} % optional
\usetikzlibrary{arrows,shapes,positioning,shadows,trees}

\newcommand{\olr}[1]{{\color{red}{}}}% \sout{#1}}}

\newcommand{\olp}[1]{{\color{cyan}{}}}% \sout{#1}}}

\newcommand{\osvp}[1]{{\color{blue}{}}}% \sout{#1}}}

\author{Sebastian V. Pios}
\affiliation{Zhejiang Laboratory, Hangzhou 311100, China}
\author{Maxim F. Gelin}
\affiliation{School of Science, Hangzhou Dianzi University, Hangzhou 310018, China}
\author{Arif Ullah}
\affiliation{School of Physics and Optoelectronic Engineering, Anhui University, Hefei 230601, China}
\author{Pavlo O. Dral}
\affiliation{State Key Laboratory of Physical Chemistry of Solid Surfaces, College of Chemistry and Chemical Engineering, Fujian Provincial Key Laboratory of Theoretical and Computational Chemistry, and Innovation Laboratory for Sciences and Technologies of Energy Materials of Fujian Province (IKKEM), Xiamen University, Xiamen, Fujian 361005, China}
\email{dral@xmu.edu.cn}
\author{Lipeng Chen}
\affiliation{Zhejiang Laboratory, Hangzhou 311100, China}
\email{chenlp@zhejianglab.com}

\title{AI-enhanced on-the-fly simulation of nonlinear time-resolved spectra}

\begin{document}

%\begin{tocentry}
%\includegraphics{toc.eps}
%\end{tocentry}

\begin{abstract}
Time-resolved spectroscopy is an important tool for unraveling the minute details of structural changes of molecules of biological and technological significance. 
The nonlinear femtosecond signals detected for such systems must be interpreted, but it is a challenging task for which theoretical simulations are often indispensable. 
Accurate simulations of transient-absorption or two-dimensional electronic spectra are, however, computationally very expensive, prohibiting the wider adoption of existing first-principles methods. 
Here, we report an AI-enhanced protocol to drastically reduce the computational cost of simulating nonlinear time-resolved electronic spectra which makes such simulations affordable for polyatomic molecules of increasing size. 
The protocol is based on doorway-window approach for the on-the-fly surface-hopping simulations. 
We show its applicability for the prototypical molecule of pyrazine for which it produces spectra with high precision with respect to \textit{ab initio} reference while cutting the computational cost by at least 95\% compared to pure first-principles simulations.

\end{abstract}

%\section{Introduction}
The rapid development of ultrafast laser techniques with capabilities to control femtosecond and even attosecond pulse durations has largely shaped our knowledge of the photoinduced reactions in polyatomic molecules. 
In particular, experiments characterizing ultrafast nonadiabatic photophysical and photochemical processes through conical intersections of electronic states of the system upon interaction with light have gained increasing attention as they are critical for a variety of applications in renewable energy, chemical synthesis, and bioimaging  
\cite{Domcke2004Book,Chen2015Review,Karkas2016CR,Krylov2017CR,Mukamel2017Review,OTF3}. 
For example, broad-band pump-probe spectroscopy \cite{Cerullo10} and   heterodyne-detected transient-grating spectroscopy \cite{DM2015NC} measurements identified key vibrational modes that drive the primary conical-intersections mediated photoisomerization reaction in rhodopsin, while transient-absorption pump-probe spectroscopy \cite{Miyata17}  and 
impulsive two- and three-pulse vibrational spectroscopy \cite{Kukura15} experiments clarified the mechanistic role of conical intersections in singlet fission in rubrene and tri-isopropylsilyly-ethynyl-pentacene (TIPS-pentacene), respectively.
Two-dimensional electronic spectroscopy \cite{Warren03,Jonas03}, which emerged as a powerful tool for monitoring energy transfer in multi-excitonic complexes  in the early 2000s  \cite{Brixner05,EET1,EET2,EET3}, was also  used to comprehensively characterize wavepacket dynamics at conical intersections, for example, in $\alpha$-terpinene \cite{Moran13}, pyrene \cite{Riedle13}, silicon 2,3-naphthalocyanine bis(trihexylsilyloxide) (SiNc) \cite{Jonas14} and DNA nucleobases \cite{Miller16}. 

However, the interpretation of time-resolved signals in terms of nonadiabatic dynamical processes is far from straightforward, and adequate theory is indispensable to decipher the intrinsic system dynamics from the detected spectroscopic responses. 
Nowadays, \textit{ab initio} simulations of spectroscopic responses are quickly establishing themselves, and a number of powerful methods have
been developed to calculate  femtosecond nonlinear spectroscopic signals of excitonic systems%%%simulations are for Frenkel excitonin models and not ab initio on the fly
\cite{Kuehn15,Mennucci18,Jansen19,Jansen21,Isborn21,Santoro21,Maxim2002CR} and
polyatomic chromophores\cite{Vanicek20a,Vanicek21,Garavelli18,Garavelli20,Garavelli21} \textit{ab initio} on-the-fly.
The majority of \textit{ab initio} on-the-fly protocols for the simulation of nonlinear spectroscopic signals hinges upon 
mixed quantum-classical approaches in which the electrons are treated quantum-mechanically and the nuclei classically \cite{OTF1,OTF2,OTF3}. 
The most widely employed mixed quantum-classical method for the description of nonadiabatic dynamics is the trajectory surface hopping method \cite{OTF2,OTF3}. 
In this method, classical trajectories are propagated on a single PES and nonadiabatic transitions between adiabatic PESs are treated via instantaneous switches, the so-called ``hops". Many algorithms have been proposed to evaluate the hopping probability, each with its advantages and disadvantages \cite{OTF1,OTF2,mai2020molecular}.
One of them, which we use in this work (see SI for details), is based on the Landau--Zener formalism\cite{LZM1,LZM2} that has an advantage of not suffering from over-coherence. 
The transition probability in this formalism can be  estimated relatively easily based on the energy difference between adjacent adiabatic electronic states and its second derivative with respect to time, thus avoiding the calculation of rapidly varying and possibly divergent nonadiabatic couplings.
One limitation introduced using the Landau--Zener formalism is the restriction that no more than two states may be close in energy for it to be reliable.\cite{pragmatic_lz}

Usually, the simulation of nonlinear time-resolved spectroscopic signals by {\it ab initio} methods requires a computationally tedious evaluation of third-order response functions \cite{Kuehn15,Mennucci18,Jansen19,Jansen21,Isborn21,Santoro21,Maxim2002CR,Vanicek20a,Vanicek21,Garavelli18,Garavelli20,Garavelli21}. However, such  signals can be extracted from the dynamical simulations using computationally much more efficient
doorway-window approach \cite{Yan89}. 
It is applicable if the laser pulses are well separated in time, i.e., are non-overlapping. The on-the-fly doorway-window simulation protocol\cite{Maxim2002CR,OTFDW1,Xiang2D} adopted in this study is based on several additional approximations, the most significant  of which is the so-called classical Condon approximation (see SI for details).

Despite the progress in developing theoretical framework for simulating time-resolved spectra with the on-the-fly approaches, their application potential is severely limited due to the high computational cost. As is becoming increasingly common in such situations, machine learning (ML) tools developed for AI research can be applied to overcome the cost limitations \cite{dral_book}. 
Successful examples in the related fields include creating surrogate ML models for excited-state properties\cite{westermayr_chemrev_2020,dral_nrc_2021}, which can be used to increase the precision of linear absorption spectra \cite{Mukamel_PNAS_2019,mlatom2}, uncertainty quantification \cite{D3CC01988H}, predict two-photon absorption cross sections \cite{mltpa}, and perform nonadiabatic molecular dynamics of molecular systems\cite{dral_deeplearning_2018,lan_ml_nadmd,westermayr_natchem_tyrosine_2022,lopez_jacs_cubanes, axelrod2022excited}. 
Beyond linear spectra, ML methods were also successfully used in the interpretation of nonlinear spectroscopic signals, i.e., for the reconstruction of certain facets of system dynamics from experimental transient-absorption pump-probe  \cite{Tisdale19, D3CP00510K} and 2D electronic spectra \cite{Li20,Fyer20,Wasilewski22,Xiong23}, as well as for ``denoising" experimental signals \cite{Baiz22}. 
A few studies invoked ML for the evaluation of nonlinear signals, i.e., for predicting 2D electronic spectra of proteins\cite{Kramer19,Markland20} and extracting relevant parameters such as orientations of transition dipole moments from these spectra \cite{Kramer19}. The application of ML for speeding up on-the-fly simulation of time-resolved spectra has remained an unexplored avenue.

In this study we introduce an approach that leverages machine learning (ML) to conduct direct, real-time trajectory simulations of the nonlinear time- and frequency-resolved spectra of polyatomic chromophores in the gas phase. 
Our adopted approach combines the well-established doorway-window protocol, used for evaluating spectroscopic signals, with ML-accelerated on-the-fly surface-hopping simulations of nonadiabatic molecular dynamics. 
To demonstrate the efficacy of our approach, we apply it to the simulation of stimulated emission contributions in transient-absorption pump-probe and two-dimensional (2D) electronic spectra of pyrazine. Pyrazine serves as a quintessential molecular system, known for its multiple conical intersections of the lower-lying excited electronic states. 
The photophysics of pyrazine has been extensively explored through \textit{ab initio} quantum-chemistry calculations~\cite{woywod_pyrazine,gatti_pyrazine}, dynamic simulations~\cite{xie_pyrazine,CLP}, femtosecond photoelectron spectroscopy measurements~\cite{radlof_pyr_phelsp,suzuki2010time,suzuki2016full}, and calculations of nonlinear spectroscopic signals~\cite{OTFDW1,Xiang2D,Skw20a,LP21,OTFDW2,OTFDW4}.

%\section{Results and discussion}
Our ML approach yields highly accurate spectra of pyrazine while significantly reducing computational costs compared to conventional (non-ML) \textit{ab initio} surface-hopping simulations. 
To achieve this, we developed a robust protocol that involves the creation of ML surrogate models which learn adiabatic state energies, forces, and oscillator strengths. 
The robustness of these models is ensured by using a well-sampled training set that covers relevant regions on PESs. 
Since it is practically impossible to guarantee that ML can confidently reproduce all regions of high-dimensional PES, particularly those near critical conical intersections, we employ uncertainty quantification to identify and address problematic areas. 
In the following, we outline the adopted protocol using pyrazine as an example, where we consider the first five singlet adiabatic states S\textsubscript{0}--S\textsubscript{4}. Our findings demonstrate that this protocol can slash the computational cost by up to 95\%.

The process of sampling training data is of utmost importance. It must strike a balance between ensuring model robustness and computational affordability. In the case of pyrazine, our training set was generated from 50 Landau--Zener surface hopping trajectories, propagated for 100~fs with a time step of 0.5~fs, using the \textit{ab initio} method ADC(2)~\cite{adc2_paper1,adc2_paper2}. 
While the number and lengths of trajectories are smaller than what is needed for precise spectra (as we will show later), they suffice to cover relevant PES regions. 
From these trajectories, we sampled 5,151 training points, capturing snapshots at approximately 1~fs intervals (50\% of the data set), while the remaining points were held back for testing. 
The training set is further divided into sub-training and validation sets in an 80:20 ratio, consisting of 4,120 and 1,031 data points, respectively.

The choice of the ML model architecture is another critical factor, and in this work, we opt for ANI-type models due to their high accuracy and efficiency in obtaining energies, gradients, and oscillator strengths within short training and prediction times (see SI for details)~\cite{barbatti_right_ml_potential}.

\begin{figure}
    \centering
    \includegraphics[width=1.0\textwidth]{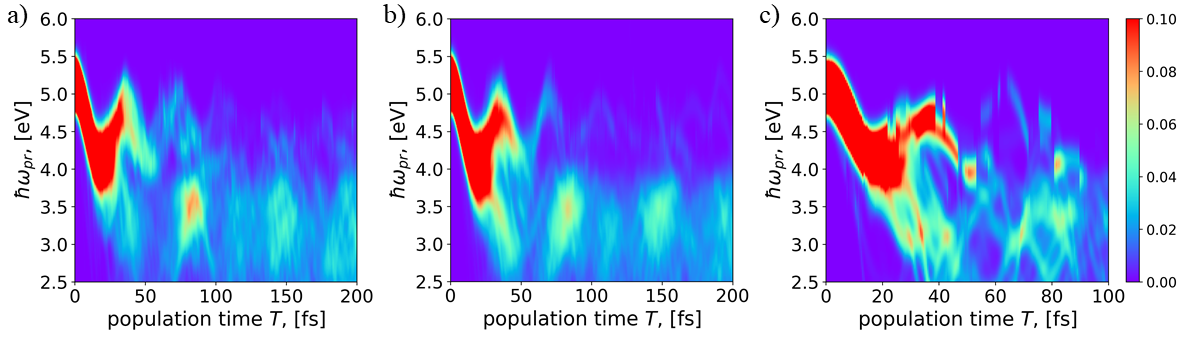}
    \caption{
    Integral time-resolved  stimulated-emission spectrum $I(T,\omega_{\mathrm{pr}})$ of pyrazine as a function of the population time $T$ and the carrier frequency $\omega_{\mathrm{pr}}$ of the probe pulse. (a). The converged spectrum calculated with 600 ML-accelerated trajectories. (b). Reference spectrum calculated 
    with 600 pure ab-initio trajectories. (c). Spectrum  calculated from 50 pure ab-initio trajectories (100~fs), which were used for the construction of the ML training set.}
    \label{fig:pumpprobe}
\end{figure}

The ML test errors for pyrazine, given in Table~\ref{error_table}, are in the range of the errors reported for slower type of ML models with more complex architecture in similar tasks \cite{westermayr_natchem_tyrosine_2022}.
The data  follow the known trend~\cite{westermayr_chemrev_2020,dral_nrc_2021} of increasing error for higher excited states. Encouragingly, the error for the oscillator strengths is also rather low with ANI-type models.
\begin{table}[]
\caption{The root-mean-square errors (RMSEs) and mean absolute errors (MAEs) in the energies {[}eV{]} and gradients {[}eV$/\text{\r{A}}${]} for their respective state as well as the RMSEs and MAEs in the oscillator strengths {[}a.u.{]} for the transition from the electronic ground-state to the corresponding excited state.}
\begin{tabular}{cccccc|ccccc}
                                          & \multicolumn{5}{c|}{RMSE} & \multicolumn{5}{c}{MAE} \\ \cline{2-11} 
                                          & GS     & S\textsubscript{1}     & S\textsubscript{2}     & S\textsubscript{3}    & S\textsubscript{4}    & GS     & S\textsubscript{1}    & S\textsubscript{2}    & S\textsubscript{3}    & S\textsubscript{4}    \\ \hline
\multicolumn{1}{c|}{Energies}             & 0.0076 & 0.0110 & 0.0139 & 0.0176 & 0.0315 & 0.0051 & 0.0082 & 0.0106 & 0.0130 & 0.0196 \\
\multicolumn{1}{c|}{Gradients}            & - &  0.0422  & 0.0833  &  0.1059  &  0.1493  & - & 0.0247 & 0.0443 & 0.0540 &  0.0833 \\ \hline
\multicolumn{1}{c|}{}                     & \multicolumn{5}{c|}{RMSE}     & \multicolumn{5}{c}{MAE}     \\ \hline
\multicolumn{1}{c|}{Osc. str.} & -      &  0.0012  &  0.0053  & 0.0068  & 0.0101  & -      & 0.0005  &  0.0022  &  0.0030  & 0.0042     
\end{tabular}
\label{error_table}
\end{table}
The accuracy of the ultrafast dynamics hinges on how well the PESs in the vicinity of conical intersections are described. 
Since these regions have a rather complex topography which is challenging for learning methods, we decided to circumvent the difficult learning problem by switching to the \textit{ab initio} method for calculating excited-state properties when the energy gap of adjacent electronic states is smaller than 0.1 eV as done previously \cite{dral_nadesmd_2018,lan_ml_nadmd}. 
In addition, we monitor the uncertainty of the ML predictions by evaluating the differences of energies and gradients generated from a pair of models for each adiabatic state, as is common for related query-by-committee strategies \cite{qbc}. 
If the deviation exceeds 0.17~eV for energies or 0.17~eV$/$\AA\space for energy gradients, we recalculate the properties for the current step with the \textit{ab initio} method.
This threshold was chosen with respect to the largest RMSE of the gradients, as especially for the S\textsubscript{4} the error increases, potentially triggering an unnecessary number of \textit{ab initio} calculations.
This mitigation strategy, depicted in Figure~\ref{fig:flowchart}, greatly enhanced the robustness of the simulations, while the number of required \textit{ab initio} calculations only slightly grew by about 1.6\%. 
We propagated an increasing number of on-the-fly dynamics trajectories for 200~fs based on the ML surrogate models with the aforementioned mitigation strategies. 
We stopped running trajectories once the resulting spectra were converged (Figure~\ref{fig:flowchart}). 
The convergence of the spectrum was judged on the mean absolute deviation of normalized intensities between the current and previous set of trajectories. Each set contains 100 more trajectories than the previous set. 
The criterion to consider the spectrum as converged was set to a mean absolute deviation of $10^{-4}$. 
For pyrazine, a total of 600 trajectories was sufficient to obtain a well-converged time-resolved stimulated emission spectrum $I(T,\omega_{\mathrm{pr}})$, which is plotted as a function of the population time $T$ and the probe-pulse  carrier frequency $\omega_{\mathrm{pr}}$ in Figure~\ref{fig:pumpprobe}a (see Figure~S2 in Supporting Information for spectrum convergence). 
Excellent agreement between the ML (Figure~\ref{fig:pumpprobe}a) and the reference {\it ab initio} (Figure~\ref{fig:pumpprobe}b) spectra in the entire time-frequency domain  demonstrates the robustness and accuracy of our approach. 
Importantly, the spectrum simulated with 600 trajectories shows a marked improvement over the spectrum obtained from 50 trajectories used to furnish the training set (Figure~\ref{fig:pumpprobe}c). 
Not just the intensities significantly improved but, remarkably, the 200-fs ML spectrum is able to accurately capture intricate features well beyond the 100-fs range of the training trajectories. 
This proves that the precision and range of the \textit{ab initio} spectrum can be greatly improved with little additional cost by ML.
\definecolor{white}{RGB}{255, 255, 255}
\definecolor{black}{RGB}{0, 0, 0}
\definecolor{grey}{RGB}{51, 51, 51}
\tikzset{
    rectanglestyle_round/.style={rectangle, rounded corners, minimum width=3cm, minimum height=1cm, text centered, font=\normalsize, color=black, draw=red, line width=1, fill=white},
    rectanglestyle/.style={rectangle, minimum width=2cm, minimum height=1cm, text centered,, font=\normalsize, color=black, draw=red, line width=1, fill=white},
    diamondstyle/.style={diamond, aspect=2.5, inner xsep=1mm, minimum width=1cm, minimum height=1cm, text width=4cm, text centered, font=\normalsize, color=black, draw=red, line width=1, top color=white, bottom color=blue!30},
    decision/.style={diamond, text centered, draw=black, aspect=5, inner xsep=1mm, top color=white, bottom color=blue!30},
    stop/.style={rectangle, rounded corners, minimum width=3cm, minimum height=1cm,text centered, draw=black},
    arrow_style/.style = {thick, draw=black, line width=2, ->, >=stealth}
    }
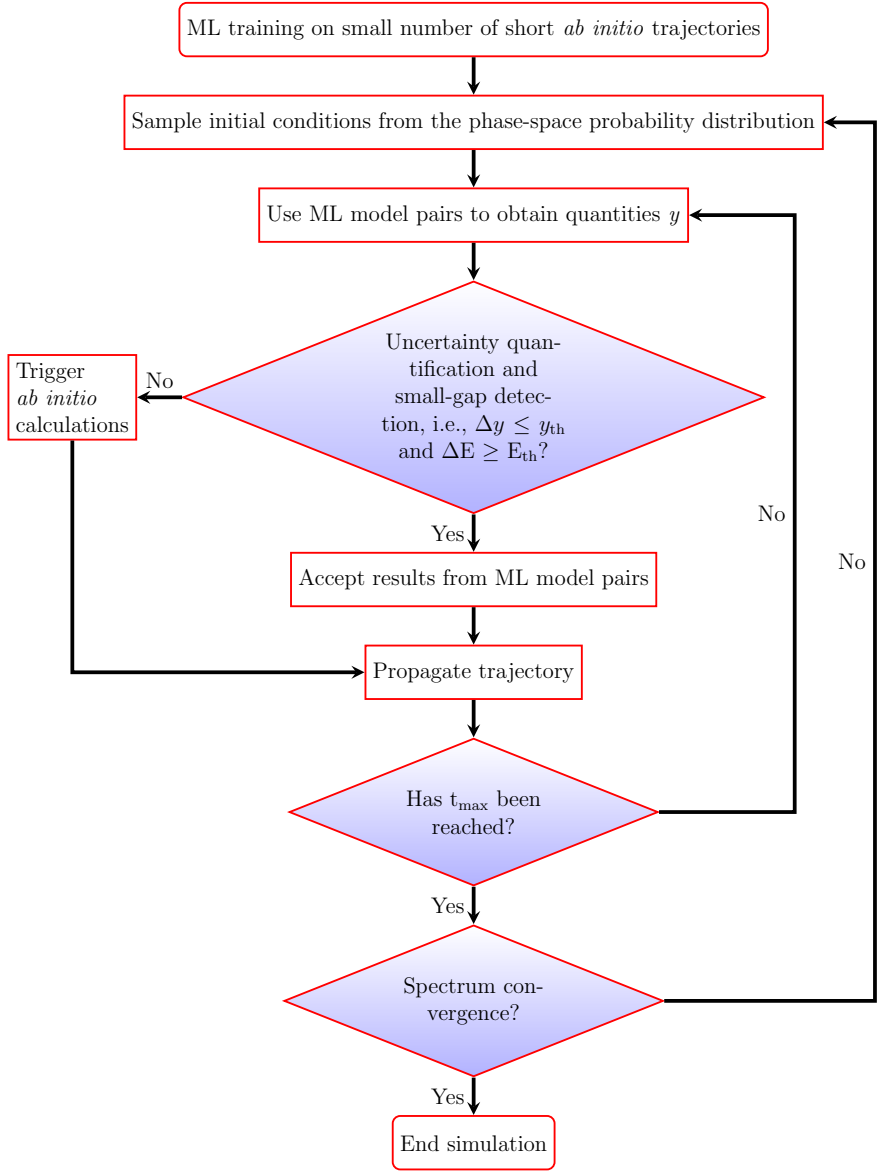
\begin{figure}[!t]
    \centering
    \hspace{0cm}
    \begin{adjustbox}{width=0.7\textwidth}
    \begin{tikzpicture}[node distance=2cm]
\node (data_generation) [rectanglestyle_round] {ML training on small number of short \textit{ab initio} trajectories};
\node (initialization) [below = 0.7cm and 3cm of data_generation,rectanglestyle] {Sample initial conditions from the phase-space probability distribution};
\node(prediction)[below = 0.7cm and 3cm of initialization, rectanglestyle]  {Use ML model pairs to obtain quantities \textit{y}};
%\node(training)[below = 1.0cm and 3cm of propagation, rectanglestyle]  {Training of ML models};
\node (mlcheck) [below=0.7cm and 3cm of prediction, diamondstyle] {Uncertainty quantification and small-gap detection, 
i.e., $\Delta y\leq y_{\rm th}$ and $\Delta$E $\geq$ E\textsubscript{th}?};
\node (triggerabinitio) [rectanglestyle, left of=mlcheck, xshift=-5.5cm, draw, align=left] {Trigger\\\textit{ab initio}\\calculations};
\node (acceptml) [below = 0.7cm and 3cm of mlcheck, rectanglestyle] {Accept results from ML model pairs};
\node (propagation) [below = 0.7cm and 3cm of acceptml, rectanglestyle] {Propagate trajectory};
\node (tcheck) [below = 0.7cm and 3cm of propagation, diamondstyle] {Has t\textsubscript{max} been reached?};
\node (convcheck) [below = 0.7cm and 3cm of tcheck, diamondstyle] {Spectrum convergence?};
\node (endsimulation) [below = 0.7cm and 3cm of convcheck, rectanglestyle_round] {End simulation};
\draw [arrow_style] (data_generation) --  (initialization);
\draw [arrow_style] (initialization) --  (prediction);
\draw [arrow_style] (prediction) --  (mlcheck);
%\draw [arrow_style] (training) --  (mlcheck);
\draw [arrow_style] (mlcheck) -- node[anchor=east] {Yes} (acceptml);
\draw [arrow_style] (mlcheck) -- node[anchor=south] {No} (triggerabinitio);
\draw [arrow_style] (acceptml) --  (propagation);
\draw [arrow_style] (triggerabinitio) |-  (propagation);
\draw [arrow_style] (propagation) --  (tcheck);
\draw [arrow_style] (tcheck) -- ++ (6,0) |- node[pos=0.25,left,anchor=east] {No} (prediction);
\draw [arrow_style] (tcheck) -- node[anchor=east] {Yes} (convcheck);
\draw [arrow_style] (convcheck)  -- ++ (7.5,0) |- node[pos=0.25,left,anchor=east] {No} (initialization);
\draw [arrow_style] (convcheck) -- node[anchor=east] {Yes} (endsimulation);
\end{tikzpicture}
\end{adjustbox}
    \caption{Flowchart depicting the steps in obtaining the time-resolved spectrum as proposed in this work. $\Delta y$ denotes the difference in the outputs of ML models in each model pair which is used for uncertainty quantification. These differences are quantified for both energies and energy gradients for each electronic state and checked whether they exceed the thresholds $y_{\rm th}$. Furthermore, we check whether the absolute energy difference between the populated and its adjacent states, $\Delta$E, exceeds the threshold E\textsubscript{th}.}
    \label{fig:flowchart}
\end{figure}

The  spectrum $I(T,\omega_{\mathrm{pr}})$ (Figures~\ref{fig:pumpprobe}a, b) mirrors the wavepacket motion in the coupled excited electronic states S\textsubscript{1}--S\textsubscript{4}, where $\omega_{\mathrm{pr}}$ monitors position(s) of the wavepacket(s) at time $T$.
This spectrum, which gives a more detailed view of the conical-intersections mediated dynamics than the excited-state adiabatic populations (see SI, Figure~S1) can be interpreted as follows. 
$I(T,\omega_{\mathrm{pr}})$ shows how a Gaussian wavepacket created by the short pump pulse in the bright B\textsubscript{2u} state ($T \approx 0$, $\hbar\omega_{\mathrm{pr}} \approx 5.2$ eV) moves ballistically driven by the slope of the PES towards the B\textsubscript{2u}--B\textsubscript{3u}  conical intersection, 
which is manifested as the bifurcation point in the spectrum around $T\approx 20$ fs and $\hbar\omega_{\mathrm{pr}} \approx 4.2$ eV.
The revival of the spectrum  at $T \approx 35$ fs and  $\hbar\omega_{\mathrm{pr}} \approx 5.0$ eV signifies the wavepacket movement to the region of its initial excitation. 
Due to the predominantly irreversible dynamics at the conical intersection, large portions of the wavepacket remain trapped in the coupled  lower-lying B\textsubscript{3u} and  A\textsubscript{u} states, which possess low oscillator strengths \cite{gatti_pyrazine} and therefore have relatively weak  contributions to the spectrum. 
$I(T,\omega_{\mathrm{pr}})$ exhibits maxima of decreasing intensity at $T\approx$ 85, 140 and 195~fs, which are caused by partial revivals of the wavepacket on the complex landscape of PESs of the coupled  B\textsubscript{2u}, B\textsubscript{3u}, and A\textsubscript{u} states. The bifurcation of the spectrum  $I(T,\omega_{\mathrm{pr}})$ at $T=20$~fs as well as its subsequent $\approx 50$~fs periodic revivals are corroborated experimentally \cite{suzuki2010time,suzuki2016full}  and by fully-quantum simulations of the models of reduced dimensionality \cite{Skw20a,LP21}.
\begin{figure}
    \centering
    \includegraphics[width=0.7\textwidth]{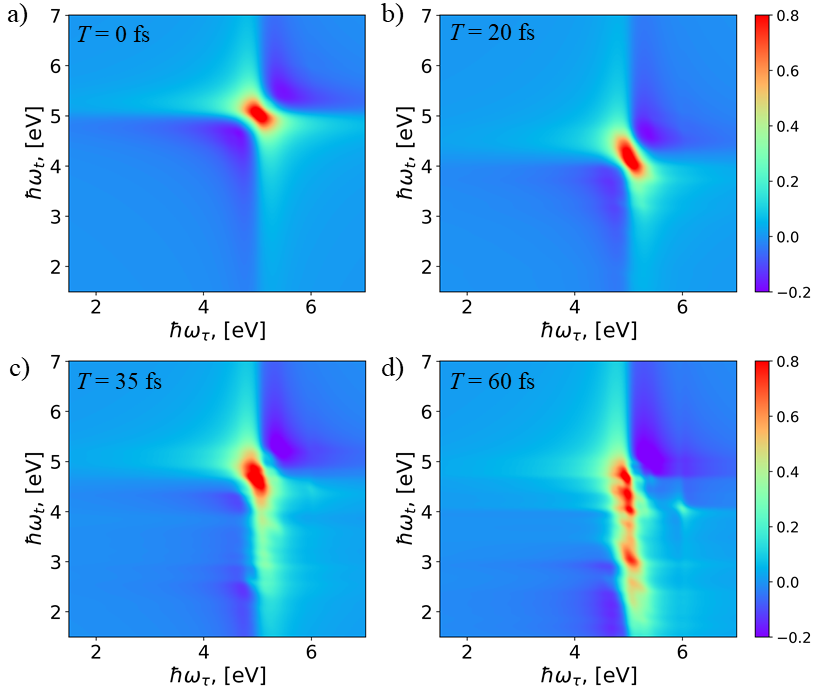}
    \caption{Non-rephasing stimulated-emission 2D electronic spectra $I_{\mathrm{2D}}(\omega_{\tau},T,\omega_{t})$ of pyrazine as a function of the excitation frequency     $\omega_{\tau}$ and the detection frequency  $\omega_{t}$  at population times $T=0$ (a), 20 (b), 35 (c), 60~fs (d) obtained with ML-accelerated doorway-window method. For each $T$, the intensities of the spectra are scaled for better visibility.}
    \label{fig:2d_evolution}
\end{figure}

The non-rephasing stimulated emission 2D spectra $I_{\mathrm{2D}}(\omega_{\tau},T,\omega_{t})$ at several population times $T$ obtained with the ML-accelerated doorway-window method are shown in Figure~\ref{fig:2d_evolution}. 
These spectra complement and refine the information on the photoinduced dynamics of pyrazine extracted from the pump-probe spectra. Namely, 2D spectra show how the wavepacket excited with photon energy $\hbar\omega_{\tau}$ and detected with photon energy $\hbar\omega_{t}$ evolves during the population time $T$. 
The single peak around $\hbar\omega_{\tau} = \hbar\omega_{t} \approx 5.2$ eV at $T=0$ fs in panel (a) is the signature of the initially-excited bright B\textsubscript{2u} state. 
The peak in panel (b) illustrates that the  wavepacket created around $\hbar\omega_{\tau}  \approx 5.2$ eV at $T=0$ fs moves to $\hbar\omega_{t}  \approx 4$ eV at $T=20$ fs, indicating the motion through the B\textsubscript{2u}--{B}\textsubscript{3u} conical intersection. 
The peak in panel (c) mirrors the partial return of the wavepacket to its initial location at $T=35$ fs. 
Panel (d) shows a progression of peaks elongated around $\hbar\omega_{t}$. These peaks reveal splitting of the wavepacket at $T=60$ fs among several areas around the flat PES minimum of the coupled A\textsubscript{u} and B\textsubscript{3u} states. 
The stimulated emission 2D spectra presented in Figure~\ref{fig:2d_evolution} are in excellent agreement with those obtained by the reference   600-trajectory surface-hopping simulations  \cite{Xiang2D} and  fully-quantum reduced-dimensional model calculations  \cite{Skw20a,LP21}. The spectrum quality is also highly dependent on the number of trajectories used (see Figure~S3 in Supporting Information).

%\section{Conclusions}
In summary, we implemented a ML-enhanced on-the-fly doorway-window protocol for the simulation of nonlinear time- and frequency-resolved spectroscopic signals. 
The protocol is  demonstrated for the computation of the stimulated-emission contribution to transient-absorption pump-probe and 2D electronic spectra, showing a drastic (95\%) reduction of the necessary \textit{ab initio} electronic structure calculations for the prototypical molecule of pyrazine. 
A number puts this into perspective: 600 trajectories comprise a total of 251,135 time steps (accounting for convergence of the hopping probability) for which calculations of quantum chemical properties have to be done. 
Our ML approach allows to reduce the number of required \textit{ab initio} calculations to only 14,352 evaluations to generate the training data and ensure the robustness of dynamics propagation. 
The switches between the predictions of the ML models and \textit{ab initio} calculations also conserved the total energy of the system, preventing nonphysical events from occurring.
The simulated stimulated emission signals of pyrazine  are in excellent agreement with the available experimental and theoretical data, which unequivocally validates the accuracy of the developed methodology. 
The latter  opens up  the possibility to investigate larger molecular systems, as the prediction time of the ML model scales linearly with the system size. 
After the inclusion of appropriate higher-lying excited electronic states, our methodology can straightforwardly be applied to the simulation of any nonlinear spectroscopic signal detected with short laser pulses and on-the-fly doorway-window protocols. 

\begin{acknowledgement}
S.V.P. and L.P.C. acknowledge support from the Key Research Project of Zhejiang Lab (No. 2021PE0AC02). M. F. G. acknowledges support from the National Natural Science Foundation of China (No.~22373028). P.O.D. acknowledges funding by the National Natural Science Foundation of China (No.~22003051) and funding via the Outstanding Youth Scholars (Overseas, 2021), the Fundamental Research Funds for the Central Universities (No.~20720210092), and via the Lab project of the State Key Laboratory of Physical Chemistry of Solid Surfaces as well as support by Science and Technology Projects of Innovation Laboratory for Sciences and Technologies of Energy Materials of Fujian Province (IKKEM) (No:~RD2022070103). The authors also acknowledge Fuchun Ge and Lina Zhang for discussions. 
\end{acknowledgement}
\section{Data availability statement}
The training and test data as well as the trained ML models are available at \url{https://github.com/psebastianzjl/pyrazine_stimulated_emission}.
\section{Code availability statement}
All modified codes from other program packages as well as the codes for predicting the stimulated emission spectra are available at \url{https://github.com/psebastianzjl/pyrazine_stimulated_emission}.
\section{Competing interests}
The authors declare that they have no competing interests.
\bibliography{main}

\providecommand{\latin}[1]{#1}
\makeatletter
\providecommand{\doi}
  {\begingroup\let\do\@makeother\dospecials
  \catcode`\{=1 \catcode`\}=2 \doi@aux}
\providecommand{\doi@aux}[1]{\endgroup\texttt{#1}}
\makeatother
\providecommand*\mcitethebibliography{\thebibliography}
\csname @ifundefined\endcsname{endmcitethebibliography}  {\let\endmcitethebibliography\endthebibliography}{}
\begin{mcitethebibliography}{79}
\providecommand*\natexlab[1]{#1}
\providecommand*\mciteSetBstSublistMode[1]{}
\providecommand*\mciteSetBstMaxWidthForm[2]{}
\providecommand*\mciteBstWouldAddEndPuncttrue
  {\def\EndOfBibitem{\unskip.}}
\providecommand*\mciteBstWouldAddEndPunctfalse
  {\let\EndOfBibitem\relax}
\providecommand*\mciteSetBstMidEndSepPunct[3]{}
\providecommand*\mciteSetBstSublistLabelBeginEnd[3]{}
\providecommand*\EndOfBibitem{}
\mciteSetBstSublistMode{f}
\mciteSetBstMaxWidthForm{subitem}{(\alph{mcitesubitemcount})}
\mciteSetBstSublistLabelBeginEnd
  {\mcitemaxwidthsubitemform\space}
  {\relax}
  {\relax}

\bibitem[Domcke \latin{et~al.}(2004)Domcke, Yarkony, and K{\"o}ppel]{Domcke2004Book}
Domcke,~W.; Yarkony,~D.~R.; K{\"o}ppel,~H. \emph{Conical Intersections: Electronic Structure, Dynamics \& Spectroscopy}; World Scientific, 2004\relax
\mciteBstWouldAddEndPuncttrue
\mciteSetBstMidEndSepPunct{\mcitedefaultmidpunct}
{\mcitedefaultendpunct}{\mcitedefaultseppunct}\relax
\EndOfBibitem
\bibitem[Chen \latin{et~al.}(2015)Chen, Shenai, Zheng, Somoza, and Zhao]{Chen2015Review}
Chen,~L.; Shenai,~P.; Zheng,~F.; Somoza,~A.; Zhao,~Y. Optimal Energy Transfer in Light-Harvesting Systems. \emph{Molecules.} \textbf{2015}, \emph{20}, 15224--15272\relax
\mciteBstWouldAddEndPuncttrue
\mciteSetBstMidEndSepPunct{\mcitedefaultmidpunct}
{\mcitedefaultendpunct}{\mcitedefaultseppunct}\relax
\EndOfBibitem
\bibitem[K\"{a}rk\"{a}s \latin{et~al.}(2016)K\"{a}rk\"{a}s, Porco~Jr, and Stephenson]{Karkas2016CR}
K\"{a}rk\"{a}s,~M.~D.; Porco~Jr,~J.~A.; Stephenson,~C. R.~J. Photochemical Approaches to Complex Chemotypes: Applications in Natural Product Synthesis. \emph{Chem. Rev.} \textbf{2016}, \emph{116}, 9683--9747\relax
\mciteBstWouldAddEndPuncttrue
\mciteSetBstMidEndSepPunct{\mcitedefaultmidpunct}
{\mcitedefaultendpunct}{\mcitedefaultseppunct}\relax
\EndOfBibitem
\bibitem[Acharya \latin{et~al.}(2017)Acharya, Bogdanov, Grigorenko, Bravaya, Nemukhin, Lukyanov, and Krylov]{Krylov2017CR}
Acharya,~A.; Bogdanov,~A.~M.; Grigorenko,~B.~L.; Bravaya,~K.~B.; Nemukhin,~A.~V.; Lukyanov,~K.~A.; Krylov,~A.~I. Photoinduced Chemistry in Fluorescent Proteins: Curse or Blessing? \emph{Chem. Rev.} \textbf{2017}, \emph{117}, 758--795\relax
\mciteBstWouldAddEndPuncttrue
\mciteSetBstMidEndSepPunct{\mcitedefaultmidpunct}
{\mcitedefaultendpunct}{\mcitedefaultseppunct}\relax
\EndOfBibitem
\bibitem[Kowalewski \latin{et~al.}(2017)Kowalewski, Fingerhut, Dorfman, Bennett, and Mukamel]{Mukamel2017Review}
Kowalewski,~M.; Fingerhut,~B.~P.; Dorfman,~K.~E.; Bennett,~K.; Mukamel,~S. Simulating Coherent Multidimensional Spectroscopy of Nonadiabatic Molecular Processes: From the Infrared to the X-ray Regime. \emph{Chem. Rev.} \textbf{2017}, \emph{117}, 12165--12226\relax
\mciteBstWouldAddEndPuncttrue
\mciteSetBstMidEndSepPunct{\mcitedefaultmidpunct}
{\mcitedefaultendpunct}{\mcitedefaultseppunct}\relax
\EndOfBibitem
\bibitem[Nelson \latin{et~al.}(2020)Nelson, White, Bjorgaard, Sifain, Zhang, Nebgen, Fernandez-Alberti, Mozyrsky, Roitberg, and Tretiak]{OTF3}
Nelson,~T.~R.; White,~A.~J.; Bjorgaard,~J.~A.; Sifain,~A.~E.; Zhang,~Y.; Nebgen,~B.; Fernandez-Alberti,~S.; Mozyrsky,~D.; Roitberg,~A.~E.; Tretiak,~S. Non-adiabatic Excited-State Molecular Dynamics: Theory and Applications for Modeling Photophysics in Extended Molecular Materials. \emph{Chem. Rev.} \textbf{2020}, \emph{120}, 2215--2287\relax
\mciteBstWouldAddEndPuncttrue
\mciteSetBstMidEndSepPunct{\mcitedefaultmidpunct}
{\mcitedefaultendpunct}{\mcitedefaultseppunct}\relax
\EndOfBibitem
\bibitem[Polli \latin{et~al.}(2010)Polli, Altoè, Weingart, Spillane, Manzoni, Brida, Tomasello, Orlandi, Kukura, Mathies, Garavelli, and Cerullo]{Cerullo10}
Polli,~D.; Altoè,~P.; Weingart,~O.; Spillane,~K.~M.; Manzoni,~C.; Brida,~D.; Tomasello,~G.; Orlandi,~G.; Kukura,~P.; Mathies,~R.~A.; Garavelli,~M.; Cerullo,~G. Conical Intersection Dynamics of the Primary Photoisomerization Event in Vision. \emph{Nature.} \textbf{2010}, \emph{467}, 440--443\relax
\mciteBstWouldAddEndPuncttrue
\mciteSetBstMidEndSepPunct{\mcitedefaultmidpunct}
{\mcitedefaultendpunct}{\mcitedefaultseppunct}\relax
\EndOfBibitem
\bibitem[Johnson \latin{et~al.}(2015)Johnson, Halpin, Morizumi, Prokhorenko, Ernst, and Miller]{DM2015NC}
Johnson,~P. J.~M.; Halpin,~A.; Morizumi,~T.; Prokhorenko,~V.~I.; Ernst,~O.~P.; Miller,~R. J.~D. Local Vibrational Coherences Drive the Primary Photochemistry of Vision. \emph{Nat. Chem.} \textbf{2015}, \emph{7}, 980--986\relax
\mciteBstWouldAddEndPuncttrue
\mciteSetBstMidEndSepPunct{\mcitedefaultmidpunct}
{\mcitedefaultendpunct}{\mcitedefaultseppunct}\relax
\EndOfBibitem
\bibitem[Miyata \latin{et~al.}(2017)Miyata, Kurashige, Watanabe, Sugimoto, Takahashi, Tanaka, Takeya, Yanai, and Matsumoto]{Miyata17}
Miyata,~K.; Kurashige,~Y.; Watanabe,~K.; Sugimoto,~T.; Takahashi,~S.; Tanaka,~S.; Takeya,~J.; Yanai,~T.; Matsumoto,~Y. Coherent Singlet Fission Activated by Symmetry Breaking. \emph{Nat. Chem.} \textbf{2017}, \emph{9}, 983--989\relax
\mciteBstWouldAddEndPuncttrue
\mciteSetBstMidEndSepPunct{\mcitedefaultmidpunct}
{\mcitedefaultendpunct}{\mcitedefaultseppunct}\relax
\EndOfBibitem
\bibitem[Musser \latin{et~al.}(2015)Musser, Liebel, Schnedermann, Wende, Kehoe, Rao, and Kukura]{Kukura15}
Musser,~A.~J.; Liebel,~M.; Schnedermann,~C.; Wende,~T.; Kehoe,~T.~B.; Rao,~A.; Kukura,~P. Evidence for Conical Intersection Dynamics Mediating Ultrafast Singlet Exciton Fission. \emph{Nat. Phys.} \textbf{2015}, \emph{11}, 352--357\relax
\mciteBstWouldAddEndPuncttrue
\mciteSetBstMidEndSepPunct{\mcitedefaultmidpunct}
{\mcitedefaultendpunct}{\mcitedefaultseppunct}\relax
\EndOfBibitem
\bibitem[Tian \latin{et~al.}(2003)Tian, Keusters, Suzaki, and Warren]{Warren03}
Tian,~P.; Keusters,~D.; Suzaki,~Y.; Warren,~W.~S. Femtosecond Phase-Coherent Two-Dimensional Spectroscopy. \emph{Science.} \textbf{2003}, \emph{300}, 1553--1555\relax
\mciteBstWouldAddEndPuncttrue
\mciteSetBstMidEndSepPunct{\mcitedefaultmidpunct}
{\mcitedefaultendpunct}{\mcitedefaultseppunct}\relax
\EndOfBibitem
\bibitem[Jonas(2003)]{Jonas03}
Jonas,~D.~M. Optical Analogs of 2D NMR. \emph{Science.} \textbf{2003}, \emph{300}, 1515--1517\relax
\mciteBstWouldAddEndPuncttrue
\mciteSetBstMidEndSepPunct{\mcitedefaultmidpunct}
{\mcitedefaultendpunct}{\mcitedefaultseppunct}\relax
\EndOfBibitem
\bibitem[Brixner \latin{et~al.}(2005)Brixner, Stenger, Vaswani, Cho, Blankenship, and Fleming]{Brixner05}
Brixner,~T.; Stenger,~J.; Vaswani,~H.~M.; Cho,~M.; Blankenship,~R.~E.; Fleming,~G.~R. Two-Dimensional Spectroscopy of Electronic Couplings in Photosynthesis. \emph{Nature.} \textbf{2005}, \emph{434}, 625--628\relax
\mciteBstWouldAddEndPuncttrue
\mciteSetBstMidEndSepPunct{\mcitedefaultmidpunct}
{\mcitedefaultendpunct}{\mcitedefaultseppunct}\relax
\EndOfBibitem
\bibitem[Thyrhaug \latin{et~al.}(2018)Thyrhaug, Tempelaar, Alcocer, \v{Z}\'{i}dek, B\'{i}na, Knoester, Jansen, and Zigmantas]{EET1}
Thyrhaug,~E.; Tempelaar,~R.; Alcocer,~M. J.~P.; \v{Z}\'{i}dek,~K.; B\'{i}na,~D.; Knoester,~J.; Jansen,~T. L.~C.; Zigmantas,~D. Identification and Characterization of Diverse Coherences in the Fenna--Matthews--Olson Complex. \emph{Nat. Chem.} \textbf{2018}, \emph{10}, 780--786\relax
\mciteBstWouldAddEndPuncttrue
\mciteSetBstMidEndSepPunct{\mcitedefaultmidpunct}
{\mcitedefaultendpunct}{\mcitedefaultseppunct}\relax
\EndOfBibitem
\bibitem[Cao \latin{et~al.}(2020)Cao, Cogdell, Coker, Duan, Hauer, Kleinekath\"{o}fer, Jansen, Man\v{c}al, Miller, Ogilvie, Prohkorenko, Renger, Tan, Tempelaar, Thorwart, Thyrhaug, Westenhoff, and Zigmantas]{EET2}
Cao,~J.~S. \latin{et~al.}  Quantum Biology Revisited. \emph{Sci. Adv.} \textbf{2020}, \emph{6}, eaaz4888\relax
\mciteBstWouldAddEndPuncttrue
\mciteSetBstMidEndSepPunct{\mcitedefaultmidpunct}
{\mcitedefaultendpunct}{\mcitedefaultseppunct}\relax
\EndOfBibitem
\bibitem[Duan \latin{et~al.}(2022)Duan, Jha, Chen, Tiwari, Cogdell, Ashraf, Prokhorenko, Thorwart, and Miller]{EET3}
Duan,~H.~G.; Jha,~A.; Chen,~L.; Tiwari,~V.; Cogdell,~R.~J.; Ashraf,~K.; Prokhorenko,~V.~I.; Thorwart,~M.; Miller,~R. J.~D. Quantum Coherent Energy Transport in the Fenna--Matthews--Olson Complex at Low Temperature. \emph{Proc. Natl. Acad. Sci. U. S. A.} \textbf{2022}, \emph{119}, e2212630119\relax
\mciteBstWouldAddEndPuncttrue
\mciteSetBstMidEndSepPunct{\mcitedefaultmidpunct}
{\mcitedefaultendpunct}{\mcitedefaultseppunct}\relax
\EndOfBibitem
\bibitem[West \latin{et~al.}(2013)West, Molesky, Montoni, and Moran]{Moran13}
West,~B.~A.; Molesky,~B.~P.; Montoni,~N.~P.; Moran,~A.~M. Nonlinear Optical Signatures of Ultraviolet Light-Induced Ring Opening in $\alpha$-Terpinene. \emph{New J. Phys.} \textbf{2013}, \emph{15}, 025007\relax
\mciteBstWouldAddEndPuncttrue
\mciteSetBstMidEndSepPunct{\mcitedefaultmidpunct}
{\mcitedefaultendpunct}{\mcitedefaultseppunct}\relax
\EndOfBibitem
\bibitem[Krebs \latin{et~al.}(2013)Krebs, Pugliesi, Hauer, and Riedle]{Riedle13}
Krebs,~N.; Pugliesi,~I.; Hauer,~J.; Riedle,~E. Two-Dimensional Fourier Transform Spectroscopy in the Ultraviolet with sub-20 fs Pump Pulses and 250-720 nm Supercontinuum Probe. \emph{New J. Phys.} \textbf{2013}, \emph{15}, 085016\relax
\mciteBstWouldAddEndPuncttrue
\mciteSetBstMidEndSepPunct{\mcitedefaultmidpunct}
{\mcitedefaultendpunct}{\mcitedefaultseppunct}\relax
\EndOfBibitem
\bibitem[Kitney-Hayes \latin{et~al.}(2014)Kitney-Hayes, Ferro, Tiwari, and Jonas]{Jonas14}
Kitney-Hayes,~K.~A.; Ferro,~A.~A.; Tiwari,~V.; Jonas,~D.~M. Two-Dimensional Fourier Transform Electronic Spectroscopy at a Conical Intersection. \emph{J. Chem. Phys.} \textbf{2014}, \emph{140}, 124312\relax
\mciteBstWouldAddEndPuncttrue
\mciteSetBstMidEndSepPunct{\mcitedefaultmidpunct}
{\mcitedefaultendpunct}{\mcitedefaultseppunct}\relax
\EndOfBibitem
\bibitem[Prokhorenko \latin{et~al.}(2016)Prokhorenko, Picchiotti, Pola, Dijkstra, and Miller]{Miller16}
Prokhorenko,~V.~I.; Picchiotti,~A.; Pola,~M.; Dijkstra,~A.~G.; Miller,~R. J.~D. New Insights into the Photophysics of DNA Nucleobases. \emph{J. Phys. Chem. Lett.} \textbf{2016}, \emph{7}, 4445--4450\relax
\mciteBstWouldAddEndPuncttrue
\mciteSetBstMidEndSepPunct{\mcitedefaultmidpunct}
{\mcitedefaultendpunct}{\mcitedefaultseppunct}\relax
\EndOfBibitem
\bibitem[Schr\"{o}ter \latin{et~al.}(2015)Schr\"{o}ter, Ivanov, Schulze, Polyutov, Yan, Pullerits, and K\"{u}hn]{Kuehn15}
Schr\"{o}ter,~M.; Ivanov,~S.~D.; Schulze,~J.; Polyutov,~S.~P.; Yan,~Y.; Pullerits,~T.; K\"{u}hn,~O. Exciton--Vibrational Coupling in the Dynamics and Spectroscopy of Frenkel Excitons in Molecular Aggregates. \emph{Phys. Rep.} \textbf{2015}, \emph{567}, 1--78\relax
\mciteBstWouldAddEndPuncttrue
\mciteSetBstMidEndSepPunct{\mcitedefaultmidpunct}
{\mcitedefaultendpunct}{\mcitedefaultseppunct}\relax
\EndOfBibitem
\bibitem[Jang and Mennucci(2018)Jang, and Mennucci]{Mennucci18}
Jang,~S.~J.; Mennucci,~B. Delocalized Excitons in Natural Light-Harvesting Complexes. \emph{Rev. Mod. Phys.} \textbf{2018}, \emph{90}, 035003\relax
\mciteBstWouldAddEndPuncttrue
\mciteSetBstMidEndSepPunct{\mcitedefaultmidpunct}
{\mcitedefaultendpunct}{\mcitedefaultseppunct}\relax
\EndOfBibitem
\bibitem[Jansen \latin{et~al.}(2019)Jansen, Saito, Jeon, and Cho]{Jansen19}
Jansen,~T. L.~C.; Saito,~S.; Jeon,~J.; Cho,~M. Theory of Coherent Two-Dimensional Vibrational Spectroscopy. \emph{J. Chem. Phys.} \textbf{2019}, \emph{150}, 100901\relax
\mciteBstWouldAddEndPuncttrue
\mciteSetBstMidEndSepPunct{\mcitedefaultmidpunct}
{\mcitedefaultendpunct}{\mcitedefaultseppunct}\relax
\EndOfBibitem
\bibitem[Jansen(2021)]{Jansen21}
Jansen,~T. L.~C. Computational Spectroscopy of Complex Systems. \emph{J. Chem. Phys.} \textbf{2021}, \emph{155}, 170901\relax
\mciteBstWouldAddEndPuncttrue
\mciteSetBstMidEndSepPunct{\mcitedefaultmidpunct}
{\mcitedefaultendpunct}{\mcitedefaultseppunct}\relax
\EndOfBibitem
\bibitem[Zuehlsdorff \latin{et~al.}(2021)Zuehlsdorff, Shedge, Lu, Hong, Aguirre, Shi, and Isborn]{Isborn21}
Zuehlsdorff,~T.~J.; Shedge,~S.~V.; Lu,~S.; Hong,~H.; Aguirre,~V.~P.; Shi,~L.; Isborn,~C.~M. Vibronic and Environmental Effects in Simulations of Optical Spectroscopy. \emph{Annu. Rev. Phys. Chem.} \textbf{2021}, \emph{72}, 165--188\relax
\mciteBstWouldAddEndPuncttrue
\mciteSetBstMidEndSepPunct{\mcitedefaultmidpunct}
{\mcitedefaultendpunct}{\mcitedefaultseppunct}\relax
\EndOfBibitem
\bibitem[Santoro \latin{et~al.}(2021)Santoro, Green, Martinez-Fernandez, Cerezo, and Improta]{Santoro21}
Santoro,~F.; Green,~J.~A.; Martinez-Fernandez,~L.; Cerezo,~J.; Improta,~R. Quantum and Semiclassical Dynamical Studies of Nonadiabatic Processes in Solution: Achievements and Perspectives. \emph{Phys. Chem. Chem. Phys.} \textbf{2021}, \emph{23}, 8181--8199\relax
\mciteBstWouldAddEndPuncttrue
\mciteSetBstMidEndSepPunct{\mcitedefaultmidpunct}
{\mcitedefaultendpunct}{\mcitedefaultseppunct}\relax
\EndOfBibitem
\bibitem[Gelin \latin{et~al.}(2022)Gelin, Chen, and Domcke]{Maxim2002CR}
Gelin,~M.~F.; Chen,~L.; Domcke,~W. Equation-of-Motion Methods for the Calculation of Femtosecond Time-Resolved 4-Wave-Mixing and N-Wave-Mixing Signals. \emph{Chem. Rev.} \textbf{2022}, \emph{122}, 17339--17396\relax
\mciteBstWouldAddEndPuncttrue
\mciteSetBstMidEndSepPunct{\mcitedefaultmidpunct}
{\mcitedefaultendpunct}{\mcitedefaultseppunct}\relax
\EndOfBibitem
\bibitem[Begu\v{s}i\'{c} and Van\'{i}\v{c}ek(2020)Begu\v{s}i\'{c}, and Van\'{i}\v{c}ek]{Vanicek20a}
Begu\v{s}i\'{c},~T.; Van\'{i}\v{c}ek,~J. On-the-Fly Ab Initio Semiclassical Evaluation of Third-Order Response Functions for Two-Dimensional Electronic Spectroscopy. \emph{J. Chem. Phys.} \textbf{2020}, \emph{153}, 184110\relax
\mciteBstWouldAddEndPuncttrue
\mciteSetBstMidEndSepPunct{\mcitedefaultmidpunct}
{\mcitedefaultendpunct}{\mcitedefaultseppunct}\relax
\EndOfBibitem
\bibitem[Begu\v{s}i\'{c} and Van\'{i}\v{c}ek(2021)Begu\v{s}i\'{c}, and Van\'{i}\v{c}ek]{Vanicek21}
Begu\v{s}i\'{c},~T.; Van\'{i}\v{c}ek,~J. Finite-Temperature, Anharmonicity, and Duschinsky Effects on the Two-Dimensional Electronic Spectra from Ab Initio Thermo-Field Gaussian Wavepacket Dynamics. \emph{J. Phys. Chem. Lett.} \textbf{2021}, \emph{12}, 2997--3005\relax
\mciteBstWouldAddEndPuncttrue
\mciteSetBstMidEndSepPunct{\mcitedefaultmidpunct}
{\mcitedefaultendpunct}{\mcitedefaultseppunct}\relax
\EndOfBibitem
\bibitem[Segarra-Mart\'{i} \latin{et~al.}(2018)Segarra-Mart\'{i}, Mukamel, Garavelli, Nenov, and Rivalta]{Garavelli18}
Segarra-Mart\'{i},~J.; Mukamel,~S.; Garavelli,~M.; Nenov,~A.; Rivalta,~I. Towards Accurate Simulation of Two-Dimensional Electronic Spectroscopy. \emph{Top. Curr. Chem.} \textbf{2018}, \emph{376}, 24\relax
\mciteBstWouldAddEndPuncttrue
\mciteSetBstMidEndSepPunct{\mcitedefaultmidpunct}
{\mcitedefaultendpunct}{\mcitedefaultseppunct}\relax
\EndOfBibitem
\bibitem[Conti \latin{et~al.}(2020)Conti, Cerullo, Nenov, and Garavelli]{Garavelli20}
Conti,~I.; Cerullo,~G.; Nenov,~A.; Garavelli,~M. Ultrafast Spectroscopy of Photoactive Molecular Systems from First Principles: Where We Stand Today and Where We Are Going. \emph{J. Am. Chem. Soc.} \textbf{2020}, \emph{142}, 16117--16139\relax
\mciteBstWouldAddEndPuncttrue
\mciteSetBstMidEndSepPunct{\mcitedefaultmidpunct}
{\mcitedefaultendpunct}{\mcitedefaultseppunct}\relax
\EndOfBibitem
\bibitem[Segatta \latin{et~al.}(2021)Segatta, Nenov, Nascimento, Govind, Mukamel, and Garavelli]{Garavelli21}
Segatta,~F.; Nenov,~A.; Nascimento,~D.~R.; Govind,~N.; Mukamel,~S.; Garavelli,~M. iSPECTRON: A Simulation Interface for Linear and Nonlinear Spectra with Ab-Initio Quantum Chemistry Software. \emph{J. Comput. Chem.} \textbf{2021}, \emph{42}, 644--659\relax
\mciteBstWouldAddEndPuncttrue
\mciteSetBstMidEndSepPunct{\mcitedefaultmidpunct}
{\mcitedefaultendpunct}{\mcitedefaultseppunct}\relax
\EndOfBibitem
\bibitem[Curchod and Mart{\'\i}nez(2018)Curchod, and Mart{\'\i}nez]{OTF1}
Curchod,~B. F.~E.; Mart{\'\i}nez,~T.~J. Ab Initio Nonadiabatic Quantum Molecular Dynamics. \emph{Chem. Rev.} \textbf{2018}, \emph{118}, 3305--3336\relax
\mciteBstWouldAddEndPuncttrue
\mciteSetBstMidEndSepPunct{\mcitedefaultmidpunct}
{\mcitedefaultendpunct}{\mcitedefaultseppunct}\relax
\EndOfBibitem
\bibitem[Crespo-Otero and Barbatti(2018)Crespo-Otero, and Barbatti]{OTF2}
Crespo-Otero,~R.; Barbatti,~M. Recent Advances and Perspectives on Nonadiabatic Mixed Quantum–Classical Dynamics. \emph{Chem. Rev.} \textbf{2018}, \emph{118}, 7026--7068\relax
\mciteBstWouldAddEndPuncttrue
\mciteSetBstMidEndSepPunct{\mcitedefaultmidpunct}
{\mcitedefaultendpunct}{\mcitedefaultseppunct}\relax
\EndOfBibitem
\bibitem[Mai and Gonz{\'a}lez(2020)Mai, and Gonz{\'a}lez]{mai2020molecular}
Mai,~S.; Gonz{\'a}lez,~L. Molecular Photochemistry: Recent Developments in Theory. \emph{Angew. Chem. Int. Ed.} \textbf{2020}, \emph{59}, 16832--16846\relax
\mciteBstWouldAddEndPuncttrue
\mciteSetBstMidEndSepPunct{\mcitedefaultmidpunct}
{\mcitedefaultendpunct}{\mcitedefaultseppunct}\relax
\EndOfBibitem
\bibitem[Belyaev \latin{et~al.}(2014)Belyaev, Lasser, and Trigila]{LZM1}
Belyaev,~A.~K.; Lasser,~C.; Trigila,~G. Landau--Zener Type Surface Hopping Algorithms. \emph{J. Chem. Phys.} \textbf{2014}, \emph{140}, 224108\relax
\mciteBstWouldAddEndPuncttrue
\mciteSetBstMidEndSepPunct{\mcitedefaultmidpunct}
{\mcitedefaultendpunct}{\mcitedefaultseppunct}\relax
\EndOfBibitem
\bibitem[Belyaev \latin{et~al.}(2015)Belyaev, Domcke, Lasser, and Trigila]{LZM2}
Belyaev,~A.~K.; Domcke,~W.; Lasser,~C.; Trigila,~G. Nonadiabatic Nuclear Dynamics of the Ammonia Cation Studied by Surface Hopping Classical Trajectory Calculations. \emph{J. Chem. Phys.} \textbf{2015}, \emph{142}, 104307\relax
\mciteBstWouldAddEndPuncttrue
\mciteSetBstMidEndSepPunct{\mcitedefaultmidpunct}
{\mcitedefaultendpunct}{\mcitedefaultseppunct}\relax
\EndOfBibitem
\bibitem[Suchan \latin{et~al.}(2020)Suchan, Janoš, and Slavíček]{pragmatic_lz}
Suchan,~J.; Janoš,~J.; Slavíček,~P. Pragmatic Approach to Photodynamics: Mixed Landau–Zener Surface Hopping with Intersystem Crossing. \emph{J. Chem. Theory Comput.} \textbf{2020}, \emph{16}, 5809--5820\relax
\mciteBstWouldAddEndPuncttrue
\mciteSetBstMidEndSepPunct{\mcitedefaultmidpunct}
{\mcitedefaultendpunct}{\mcitedefaultseppunct}\relax
\EndOfBibitem
\bibitem[Yan \latin{et~al.}(1989)Yan, Fried, and Mukamel]{Yan89}
Yan,~Y.~J.; Fried,~L.~E.; Mukamel,~S. Ultrafast Pump-Probe Spectroscopy: Femtosecond Dynamics in Liouville Space. \emph{J. Phys. Chem.} \textbf{1989}, \emph{93}, 8149--8162\relax
\mciteBstWouldAddEndPuncttrue
\mciteSetBstMidEndSepPunct{\mcitedefaultmidpunct}
{\mcitedefaultendpunct}{\mcitedefaultseppunct}\relax
\EndOfBibitem
\bibitem[Gelin \latin{et~al.}(2021)Gelin, Huang, Xie, Chen, Do{\v{s}}li{\'c}, and Domcke]{OTFDW1}
Gelin,~M.~F.; Huang,~X.; Xie,~W.; Chen,~L.; Do{\v{s}}li{\'c},~N.; Domcke,~W. Ab Initio Surface-Hopping Simulation of Femtosecond Transient-Absorption Pump–Probe Signals of Nonadiabatic Excited-State Dynamics Using the Doorway–Window Representation. \emph{J. Chem. Theory Comput.} \textbf{2021}, \emph{17}, 2394--2408\relax
\mciteBstWouldAddEndPuncttrue
\mciteSetBstMidEndSepPunct{\mcitedefaultmidpunct}
{\mcitedefaultendpunct}{\mcitedefaultseppunct}\relax
\EndOfBibitem
\bibitem[Huang \latin{et~al.}(2021)Huang, Xie, Do{\v{s}}li{\'c}, Gelin, and Domcke]{Xiang2D}
Huang,~X.; Xie,~W.; Do{\v{s}}li{\'c},~N.; Gelin,~M.~F.; Domcke,~W. Ab Initio Quasiclassical Simulation of Femtosecond Time-Resolved Two-Dimensional Electronic Spectra of Pyrazine. \emph{J. Phys. Chem. Lett.} \textbf{2021}, \emph{12}, 11736--11744\relax
\mciteBstWouldAddEndPuncttrue
\mciteSetBstMidEndSepPunct{\mcitedefaultmidpunct}
{\mcitedefaultendpunct}{\mcitedefaultseppunct}\relax
\EndOfBibitem
\bibitem[Dral(2022)]{dral_book}
Dral,~P.~O. \emph{Quantum Chemistry in the Age of Machine Learning}; Elsevier, 2022\relax
\mciteBstWouldAddEndPuncttrue
\mciteSetBstMidEndSepPunct{\mcitedefaultmidpunct}
{\mcitedefaultendpunct}{\mcitedefaultseppunct}\relax
\EndOfBibitem
\bibitem[Westermayr and Marquetand(2021)Westermayr, and Marquetand]{westermayr_chemrev_2020}
Westermayr,~J.; Marquetand,~P. Machine Learning for Electronically Excited States of Molecules. \emph{Chem. Rev.} \textbf{2021}, \emph{121}, 9873--9926\relax
\mciteBstWouldAddEndPuncttrue
\mciteSetBstMidEndSepPunct{\mcitedefaultmidpunct}
{\mcitedefaultendpunct}{\mcitedefaultseppunct}\relax
\EndOfBibitem
\bibitem[Dral and Barbatti(2021)Dral, and Barbatti]{dral_nrc_2021}
Dral,~P.~O.; Barbatti,~M. Molecular Excited States through a Machine Learning Lens. \emph{Nat. Rev. Chem.} \textbf{2021}, \emph{5}, 388--405\relax
\mciteBstWouldAddEndPuncttrue
\mciteSetBstMidEndSepPunct{\mcitedefaultmidpunct}
{\mcitedefaultendpunct}{\mcitedefaultseppunct}\relax
\EndOfBibitem
\bibitem[Ye \latin{et~al.}(2019)Ye, Hu, Li, Zhang, Zhong, Zhang, Luo, Mukamel, and Jiang]{Mukamel_PNAS_2019}
Ye,~S.; Hu,~W.; Li,~X.; Zhang,~J.; Zhong,~K.; Zhang,~G.; Luo,~Y.; Mukamel,~S.; Jiang,~J. A Neural Network Protocol for Electronic Excitations of N-Methylacetamide. \emph{Proc. Natl. Acad. Sci. U.S.A.} \textbf{2019}, \emph{116}, 11612--11617\relax
\mciteBstWouldAddEndPuncttrue
\mciteSetBstMidEndSepPunct{\mcitedefaultmidpunct}
{\mcitedefaultendpunct}{\mcitedefaultseppunct}\relax
\EndOfBibitem
\bibitem[Dral \latin{et~al.}(2021)Dral, Ge, Xue, Hou, Pinheiro~Jr, Huang, and Barbatti]{mlatom2}
Dral,~P.~O.; Ge,~F.; Xue,~B.-X.; Hou,~Y.-F.; Pinheiro~Jr,~M.; Huang,~J.; Barbatti,~M. MLatom 2: An Integrative Platform for Atomistic Machine Learning. \emph{Top. Curr. Chem.} \textbf{2021}, \emph{379}, 27\relax
\mciteBstWouldAddEndPuncttrue
\mciteSetBstMidEndSepPunct{\mcitedefaultmidpunct}
{\mcitedefaultendpunct}{\mcitedefaultseppunct}\relax
\EndOfBibitem
\bibitem[Verma \latin{et~al.}(2023)Verma, Aznan, Garside, and Penfold]{D3CC01988H}
Verma,~S.; Aznan,~N. K.~N.; Garside,~K.; Penfold,~T.~J. Uncertainty quantification of spectral predictions using deep neural networks. \emph{Chem. Commun.} \textbf{2023}, \emph{59}, 7100--7103\relax
\mciteBstWouldAddEndPuncttrue
\mciteSetBstMidEndSepPunct{\mcitedefaultmidpunct}
{\mcitedefaultendpunct}{\mcitedefaultseppunct}\relax
\EndOfBibitem
\bibitem[Su \latin{et~al.}(2023)Su, Dai, Zeng, Wei, Chen, Ge, Zheng, Zhou, Dral, and Wang]{mltpa}
Su,~Y.; Dai,~Y.; Zeng,~Y.; Wei,~C.; Chen,~Y.; Ge,~F.; Zheng,~P.; Zhou,~D.; Dral,~P.~O.; Wang,~C. Interpretable Machine Learning of Two-Photon Absorption. \emph{Adv. Sci.} \textbf{2023}, \emph{10}, 2204902\relax
\mciteBstWouldAddEndPuncttrue
\mciteSetBstMidEndSepPunct{\mcitedefaultmidpunct}
{\mcitedefaultendpunct}{\mcitedefaultseppunct}\relax
\EndOfBibitem
\bibitem[Chen \latin{et~al.}(2018)Chen, Liu, Fang, Dral, and Cui]{dral_deeplearning_2018}
Chen,~W.-K.; Liu,~X.-Y.; Fang,~W.-H.; Dral,~P.~O.; Cui,~G. Deep Learning for Nonadiabatic Excited-State Dynamics. \emph{J. Phys. Chem. Lett.} \textbf{2018}, \emph{9}, 6702--6708\relax
\mciteBstWouldAddEndPuncttrue
\mciteSetBstMidEndSepPunct{\mcitedefaultmidpunct}
{\mcitedefaultendpunct}{\mcitedefaultseppunct}\relax
\EndOfBibitem
\bibitem[Hu \latin{et~al.}(2018)Hu, Xie, Li, Li, and Lan]{lan_ml_nadmd}
Hu,~D.; Xie,~Y.; Li,~X.; Li,~L.; Lan,~Z. Inclusion of Machine Learning Kernel Ridge Regression Potential Energy Surfaces in On-the-Fly Nonadiabatic Molecular Dynamics Simulation. \emph{J. Phys. Chem. Lett.} \textbf{2018}, \emph{9}, 2725--2732\relax
\mciteBstWouldAddEndPuncttrue
\mciteSetBstMidEndSepPunct{\mcitedefaultmidpunct}
{\mcitedefaultendpunct}{\mcitedefaultseppunct}\relax
\EndOfBibitem
\bibitem[Westermayr \latin{et~al.}(2022)Westermayr, Gastegger, V{\"o}r{\"o}s, Panzenboeck, Joerg, Gonz{\'a}lez, and Marquetand]{westermayr_natchem_tyrosine_2022}
Westermayr,~J.; Gastegger,~M.; V{\"o}r{\"o}s,~D.; Panzenboeck,~L.; Joerg,~F.; Gonz{\'a}lez,~L.; Marquetand,~P. Deep Learning Study of Tyrosine Reveals that Roaming can Lead to Photodamage. \emph{Nat. Chem.} \textbf{2022}, \emph{14}, 914--919\relax
\mciteBstWouldAddEndPuncttrue
\mciteSetBstMidEndSepPunct{\mcitedefaultmidpunct}
{\mcitedefaultendpunct}{\mcitedefaultseppunct}\relax
\EndOfBibitem
\bibitem[Li \latin{et~al.}(2021)Li, Stein, Adrion, and Lopez]{lopez_jacs_cubanes}
Li,~J.; Stein,~R.; Adrion,~D.~M.; Lopez,~S.~A. Machine-Learning Photodynamics Simulations Uncover the Role of Substituent Effects on the Photochemical Formation of Cubanes. \emph{J. Am. Chem. Soc.} \textbf{2021}, \emph{143}, 20166--20175\relax
\mciteBstWouldAddEndPuncttrue
\mciteSetBstMidEndSepPunct{\mcitedefaultmidpunct}
{\mcitedefaultendpunct}{\mcitedefaultseppunct}\relax
\EndOfBibitem
\bibitem[Axelrod \latin{et~al.}(2022)Axelrod, Shakhnovich, and G{\'o}mez-Bombarelli]{axelrod2022excited}
Axelrod,~S.; Shakhnovich,~E.; G{\'o}mez-Bombarelli,~R. Excited state non-adiabatic dynamics of large photoswitchable molecules using a chemically transferable machine learning potential. \emph{Nat. Commun.} \textbf{2022}, \emph{13}, 3440\relax
\mciteBstWouldAddEndPuncttrue
\mciteSetBstMidEndSepPunct{\mcitedefaultmidpunct}
{\mcitedefaultendpunct}{\mcitedefaultseppunct}\relax
\EndOfBibitem
\bibitem[Ashner \latin{et~al.}(2019)Ashner, Winslow, Swan, and Tisdale]{Tisdale19}
Ashner,~M.~N.; Winslow,~S.~W.; Swan,~J.~W.; Tisdale,~W.~A. Markov Chain Monte Carlo Sampling for Target Analysis of Transient Absorption Spectra. \emph{J. Phys. Chem. A.} \textbf{2019}, \emph{123}, 3893--3902\relax
\mciteBstWouldAddEndPuncttrue
\mciteSetBstMidEndSepPunct{\mcitedefaultmidpunct}
{\mcitedefaultendpunct}{\mcitedefaultseppunct}\relax
\EndOfBibitem
\bibitem[Middleton \latin{et~al.}(2023)Middleton, Rankine, and Penfold]{D3CP00510K}
Middleton,~C.; Rankine,~C.~D.; Penfold,~T.~J. An on-the-fly deep neural network for simulating time-resolved spectroscopy: predicting the ultrafast ring opening dynamics of 1{,}2-dithiane. \emph{Phys. Chem. Chem. Phys.} \textbf{2023}, \emph{25}, 13325--13334\relax
\mciteBstWouldAddEndPuncttrue
\mciteSetBstMidEndSepPunct{\mcitedefaultmidpunct}
{\mcitedefaultendpunct}{\mcitedefaultseppunct}\relax
\EndOfBibitem
\bibitem[Namuduri \latin{et~al.}(2020)Namuduri, Titze, Bhansali, and Li]{Li20}
Namuduri,~S.; Titze,~M.; Bhansali,~S.; Li,~H. Machine Learning Enabled Lineshape Analysis in Optical Two-Dimensional Coherent Spectroscopy. \emph{J. Opt. Soc. Am. B.} \textbf{2020}, \emph{37}, 1587--1591\relax
\mciteBstWouldAddEndPuncttrue
\mciteSetBstMidEndSepPunct{\mcitedefaultmidpunct}
{\mcitedefaultendpunct}{\mcitedefaultseppunct}\relax
\EndOfBibitem
\bibitem[Hoffman and Fayer(2020)Hoffman, and Fayer]{Fyer20}
Hoffman,~D.~J.; Fayer,~M.~D. CLS Next Gen: Accurate Frequency–Frequency Correlation Functions from Center Line Slope Analysis of 2D Correlation Spectra Using Artificial Neural Networks. \emph{J. Phys. Chem. A.} \textbf{2020}, \emph{124}, 5979--5992\relax
\mciteBstWouldAddEndPuncttrue
\mciteSetBstMidEndSepPunct{\mcitedefaultmidpunct}
{\mcitedefaultendpunct}{\mcitedefaultseppunct}\relax
\EndOfBibitem
\bibitem[Parker \latin{et~al.}(2022)Parker, Schultz, Singh, Wasielewski, and Beratan]{Wasilewski22}
Parker,~K.~A.; Schultz,~J.~D.; Singh,~N.; Wasielewski,~M.~R.; Beratan,~D.~N. Mapping Simulated Two-Dimensional Spectra to Molecular Models Using Machine Learning. \emph{J. Phys. Chem. Lett.} \textbf{2022}, \emph{13}, 7454--7461\relax
\mciteBstWouldAddEndPuncttrue
\mciteSetBstMidEndSepPunct{\mcitedefaultmidpunct}
{\mcitedefaultendpunct}{\mcitedefaultseppunct}\relax
\EndOfBibitem
\bibitem[Valentine \latin{et~al.}(2023)Valentine, Wiesehan, and Xiong]{Xiong23}
Valentine,~M.~L.; Wiesehan,~G.~D.; Xiong,~W. An Evaluation of Maximum Determination Methods for Center Line Slope Analysis. \emph{J. Phys. Chem. B.} \textbf{2023}, \emph{127}, 4268–4276\relax
\mciteBstWouldAddEndPuncttrue
\mciteSetBstMidEndSepPunct{\mcitedefaultmidpunct}
{\mcitedefaultendpunct}{\mcitedefaultseppunct}\relax
\EndOfBibitem
\bibitem[Al-Mualem and Baiz(2022)Al-Mualem, and Baiz]{Baiz22}
Al-Mualem,~Z.~A.; Baiz,~C.~R. Generative Adversarial Neural Networks for Denoising Coherent Multidimensional Spectra. \emph{J. Phys. Chem. A.} \textbf{2022}, \emph{126}, 3816--3825\relax
\mciteBstWouldAddEndPuncttrue
\mciteSetBstMidEndSepPunct{\mcitedefaultmidpunct}
{\mcitedefaultendpunct}{\mcitedefaultseppunct}\relax
\EndOfBibitem
\bibitem[Rodr{\'\i}guez and Kramer(2019)Rodr{\'\i}guez, and Kramer]{Kramer19}
Rodr{\'\i}guez,~M.; Kramer,~T. Machine Learning of Two-Dimensional Spectroscopic Data. \emph{Chem. Phys.} \textbf{2019}, \emph{520}, 52--60\relax
\mciteBstWouldAddEndPuncttrue
\mciteSetBstMidEndSepPunct{\mcitedefaultmidpunct}
{\mcitedefaultendpunct}{\mcitedefaultseppunct}\relax
\EndOfBibitem
\bibitem[Chen \latin{et~al.}(2020)Chen, Zuehlsdorff, Morawietz, Isborn, and Markland]{Markland20}
Chen,~M.~S.; Zuehlsdorff,~T.~J.; Morawietz,~T.; Isborn,~C.~M.; Markland,~T.~E. Exploiting Machine Learning to Efficiently Predict Multidimensional Optical Spectra in Complex Environments. \emph{J. Phys. Chem. Lett.} \textbf{2020}, \emph{11}, 7559--7568\relax
\mciteBstWouldAddEndPuncttrue
\mciteSetBstMidEndSepPunct{\mcitedefaultmidpunct}
{\mcitedefaultendpunct}{\mcitedefaultseppunct}\relax
\EndOfBibitem
\bibitem[Woywod \latin{et~al.}(1994)Woywod, Domcke, Sobolewski, and Werner]{woywod_pyrazine}
Woywod,~C.; Domcke,~W.; Sobolewski,~A.~L.; Werner,~H.-J. Characterization of the $\mathrm{S_1}$--$\mathrm{S_2}$ Conical Intersection in Pyrazine Using \textit{ab initio} Multiconfiguration Self-Consistent-Field and Multireference Configuration-Interaction Methods. \emph{J. Chem. Phys.} \textbf{1994}, \emph{100}, 1400--1413\relax
\mciteBstWouldAddEndPuncttrue
\mciteSetBstMidEndSepPunct{\mcitedefaultmidpunct}
{\mcitedefaultendpunct}{\mcitedefaultseppunct}\relax
\EndOfBibitem
\bibitem[Sala \latin{et~al.}(2014)Sala, Lasorne, Gatti, and Gu{\'e}rin]{gatti_pyrazine}
Sala,~M.; Lasorne,~B.; Gatti,~F.; Gu{\'e}rin,~S. The Role of the Low-lying Dark n$\pi$* States in the Photophysics of Pyrazine: a Quantum Dynamics Study. \emph{Phys. Chem. Chem. Phys.} \textbf{2014}, \emph{16}, 15957--15967\relax
\mciteBstWouldAddEndPuncttrue
\mciteSetBstMidEndSepPunct{\mcitedefaultmidpunct}
{\mcitedefaultendpunct}{\mcitedefaultseppunct}\relax
\EndOfBibitem
\bibitem[Xie \latin{et~al.}(2019)Xie, Sapunar, Do{\v{s}}li{\'c}, Sala, and Domcke]{xie_pyrazine}
Xie,~W.; Sapunar,~M.; Do{\v{s}}li{\'c},~N.; Sala,~M.; Domcke,~W. Assessing the Performance of Trajectory Surface Hopping Methods: Ultrafast Internal Conversion in Pyrazine. \emph{J. Chem. Phys.} \textbf{2019}, \emph{150}, 154119\relax
\mciteBstWouldAddEndPuncttrue
\mciteSetBstMidEndSepPunct{\mcitedefaultmidpunct}
{\mcitedefaultendpunct}{\mcitedefaultseppunct}\relax
\EndOfBibitem
\bibitem[Chen \latin{et~al.}(2019)Chen, Gelin, and Domcke]{CLP}
Chen,~L.; Gelin,~M.~F.; Domcke,~W. Multimode Quantum Dynamics with Multiple Davydov D2 Trial States: Application to a 24-Dimensional Conical Intersection Model. \emph{J. Chem. Phys.} \textbf{2019}, \emph{150}, 024101\relax
\mciteBstWouldAddEndPuncttrue
\mciteSetBstMidEndSepPunct{\mcitedefaultmidpunct}
{\mcitedefaultendpunct}{\mcitedefaultseppunct}\relax
\EndOfBibitem
\bibitem[Stert \latin{et~al.}(2000)Stert, Farmanara, and Radloff]{radlof_pyr_phelsp}
Stert,~V.; Farmanara,~P.; Radloff,~W. Electron Configuration Changes in Excited Pyrazine Molecules Analyzed by Femtosecond Time-Resolved Photoelectron Spectroscopy. \emph{J. Chem. Phys.} \textbf{2000}, \emph{112}, 4460--4464\relax
\mciteBstWouldAddEndPuncttrue
\mciteSetBstMidEndSepPunct{\mcitedefaultmidpunct}
{\mcitedefaultendpunct}{\mcitedefaultseppunct}\relax
\EndOfBibitem
\bibitem[Suzuki \latin{et~al.}(2010)Suzuki, Fuji, Horio, and Suzuki]{suzuki2010time}
Suzuki,~Y.~I.; Fuji,~T.; Horio,~T.; Suzuki,~T. Time-Resolved Photoelectron Imaging of Ultrafast S\textsubscript{2}→ S\textsubscript{1} Internal Conversion through Conical Intersection in Pyrazine. \emph{J. Chem. Phys.} \textbf{2010}, \emph{132}, 174302\relax
\mciteBstWouldAddEndPuncttrue
\mciteSetBstMidEndSepPunct{\mcitedefaultmidpunct}
{\mcitedefaultendpunct}{\mcitedefaultseppunct}\relax
\EndOfBibitem
\bibitem[Horio \latin{et~al.}(2016)Horio, Spesyvtsev, Nagashima, Ingle, Suzuki, and Suzuki]{suzuki2016full}
Horio,~T.; Spesyvtsev,~R.; Nagashima,~K.; Ingle,~R.~A.; Suzuki,~Y.~I.; Suzuki,~T. Full Observation of Ultrafast Cascaded Radiationless Transitions from S\textsubscript{2} ($\pi{\pi}^{*}$) State of Pyrazine using Vacuum Ultraviolet Photoelectron Imaging. \emph{J. Chem. Phys.} \textbf{2016}, \emph{145}, 044306\relax
\mciteBstWouldAddEndPuncttrue
\mciteSetBstMidEndSepPunct{\mcitedefaultmidpunct}
{\mcitedefaultendpunct}{\mcitedefaultseppunct}\relax
\EndOfBibitem
\bibitem[Sun \latin{et~al.}(2020)Sun, Xie, Chen, Domcke, and Gelin]{Skw20a}
Sun,~K.; Xie,~W.; Chen,~L.; Domcke,~W.; Gelin,~M.~F. Multi-Faceted Spectroscopic Mapping of Ultrafast Nonadiabatic Dynamics near Conical Intersections: A Computational Study. \emph{J. Chem. Phys.} \textbf{2020}, \emph{153}, 174111\relax
\mciteBstWouldAddEndPuncttrue
\mciteSetBstMidEndSepPunct{\mcitedefaultmidpunct}
{\mcitedefaultendpunct}{\mcitedefaultseppunct}\relax
\EndOfBibitem
\bibitem[Chen \latin{et~al.}(2021)Chen, Sun, Shalashilin, Gelin, and Zhao]{LP21}
Chen,~L.; Sun,~K.; Shalashilin,~D.~V.; Gelin,~M.~F.; Zhao,~Y. Efficient Simulation of Time- and Frequency-Resolved Four-Wave-Mixing Signals with a Multiconfigurational Ehrenfest Approach. \emph{J. Chem. Phys.} \textbf{2021}, \emph{154}, 054105\relax
\mciteBstWouldAddEndPuncttrue
\mciteSetBstMidEndSepPunct{\mcitedefaultmidpunct}
{\mcitedefaultendpunct}{\mcitedefaultseppunct}\relax
\EndOfBibitem
\bibitem[Pites{\v{s}}{\v{a}} \latin{et~al.}(2021)Pites{\v{s}}{\v{a}}, Sapunar, Ponzi, Gelin, Do{\v{s}}li{\'c}, Domcke, and Decleva]{OTFDW2}
Pites{\v{s}}{\v{a}},~T.; Sapunar,~M.; Ponzi,~A.; Gelin,~M.~F.; Do{\v{s}}li{\'c},~N.; Domcke,~W.; Decleva,~P. Combined Surface-Hopping, Dyson Orbital, and B-Spline Approach for the Computation of Time-Resolved Photoelectron Spectroscopy Signals: The Internal Conversion in Pyrazine. \emph{J. Chem. Theory Comput.} \textbf{2021}, \emph{17}, 5098--5109\relax
\mciteBstWouldAddEndPuncttrue
\mciteSetBstMidEndSepPunct{\mcitedefaultmidpunct}
{\mcitedefaultendpunct}{\mcitedefaultseppunct}\relax
\EndOfBibitem
\bibitem[Kaczun \latin{et~al.}(2023)Kaczun, Dempwolff, Huang, Gelin, Domcke, and Dreuw]{OTFDW4}
Kaczun,~T.; Dempwolff,~A.~L.; Huang,~X.; Gelin,~M.~F.; Domcke,~W.; Dreuw,~A. Tuning UV Pump X-ray Probe Spectroscopy on the Nitrogen K Edge Reveals the Radiationless Relaxation of Pyrazine: Ab Initio Simulations Using the Quasiclassical Doorway--Window Approximation. \emph{J. Phys. Chem. Lett.} \textbf{2023}, \emph{14}, 5648--5656\relax
\mciteBstWouldAddEndPuncttrue
\mciteSetBstMidEndSepPunct{\mcitedefaultmidpunct}
{\mcitedefaultendpunct}{\mcitedefaultseppunct}\relax
\EndOfBibitem
\bibitem[Trofimov and Schirmer(1995)Trofimov, and Schirmer]{adc2_paper1}
Trofimov,~A.~B.; Schirmer,~J. An Efficient Polarization Propagator Approach to Valence Electron Excitation Spectra. \emph{J. Phys. B: At. Mol. Opt. Phys.} \textbf{1995}, \emph{28}, 2299\relax
\mciteBstWouldAddEndPuncttrue
\mciteSetBstMidEndSepPunct{\mcitedefaultmidpunct}
{\mcitedefaultendpunct}{\mcitedefaultseppunct}\relax
\EndOfBibitem
\bibitem[Dreuw and Wormit(2015)Dreuw, and Wormit]{adc2_paper2}
Dreuw,~A.; Wormit,~M. The Algebraic Diagrammatic Construction Scheme for the Polarization Propagator for the Calculation of Excited States. \emph{Wiley Interdiscip. Rev. Comput. Mol. Sci.} \textbf{2015}, \emph{5}, 82--95\relax
\mciteBstWouldAddEndPuncttrue
\mciteSetBstMidEndSepPunct{\mcitedefaultmidpunct}
{\mcitedefaultendpunct}{\mcitedefaultseppunct}\relax
\EndOfBibitem
\bibitem[Pinheiro \latin{et~al.}(2021)Pinheiro, Ge, Ferr{\'e}, Dral, and Barbatti]{barbatti_right_ml_potential}
Pinheiro,~M.; Ge,~F.; Ferr{\'e},~N.; Dral,~P.~O.; Barbatti,~M. Choosing the Right Molecular Machine Learning Potential. \emph{Chem. Sci.} \textbf{2021}, \emph{12}, 14396--14413\relax
\mciteBstWouldAddEndPuncttrue
\mciteSetBstMidEndSepPunct{\mcitedefaultmidpunct}
{\mcitedefaultendpunct}{\mcitedefaultseppunct}\relax
\EndOfBibitem
\bibitem[Dral \latin{et~al.}(2018)Dral, Barbatti, and Thiel]{dral_nadesmd_2018}
Dral,~P.~O.; Barbatti,~M.; Thiel,~W. Nonadiabatic Excited-State Dynamics with Machine Learning. \emph{J. Phys. Chem. Lett.} \textbf{2018}, \emph{9}, 5660--5663\relax
\mciteBstWouldAddEndPuncttrue
\mciteSetBstMidEndSepPunct{\mcitedefaultmidpunct}
{\mcitedefaultendpunct}{\mcitedefaultseppunct}\relax
\EndOfBibitem
\bibitem[Freund \latin{et~al.}(1997)Freund, Seung, Shamir, and Tishby]{qbc}
Freund,~Y.; Seung,~H.~S.; Shamir,~E.; Tishby,~N. Selective Sampling using the Query by Committee Algorithm. \emph{Mach. Learn.} \textbf{1997}, \emph{28}, 133--168\relax
\mciteBstWouldAddEndPuncttrue
\mciteSetBstMidEndSepPunct{\mcitedefaultmidpunct}
{\mcitedefaultendpunct}{\mcitedefaultseppunct}\relax
\EndOfBibitem
\end{mcitethebibliography}
\end{document}